\documentclass[preprint2]{aastex}
\input psfig.sty

\newcommand{\nh}{N_{\rm H}}
\newcommand{\sax}{{\it Beppo\-SAX}}
\newcommand{\source}{XTE\ J1118+480}
\newcommand{\msun}{{\rm M}_\sun}
\newcommand{\xte}{{\it RXTE}}

\newcommand{\asm}{{\it RXTE}/ASM}

\newcommand{\ergcms}{\mbox{erg cm$^{-2}$ s$^{-1}$}}

\begin{document}

\title{A measurement of the broad-band spectrum of XTE J1118+480 with BeppoSAX
and its astrophysical implications}

\author{F. Frontera\altaffilmark{1,2},
A. A.~Zdziarski\altaffilmark{3},
L. Amati\altaffilmark{2},
J. Miko{\l}ajewska\altaffilmark{3},
T. Belloni\altaffilmark{4},
%D.~Dal Fiume\altaffilmark{1},
S.~Del~Sordo\altaffilmark{5},
F.~Haardt\altaffilmark{6},
E. Kuulkers\altaffilmark{7},
N. Masetti\altaffilmark{2},
M.~Orlandini\altaffilmark{2},
E. Palazzi\altaffilmark{2},
A. N.~Parmar\altaffilmark{8},
R. A. Remillard\altaffilmark{9},
A.~Santangelo\altaffilmark{5},
L. Stella\altaffilmark{10}
}
\altaffiltext{1}{Dipartimento di Fisica, Universit\`a degli Studi di Ferrara,
Via Paradiso 12, I-44100 Ferrara, Italy; frontera@fe.infn.it}

\altaffiltext{2}{Istituto Tecnologie e Studio Radiazioni Extraterrestri, CNR,
Via Gobetti 101, I-40129 Bologna, Italy}

\altaffiltext{3}{N. Copernicus Astronomical Center, Bartycka 18,
00-716 Warsaw, Poland; aaz@camk.edu.pl}

\altaffiltext{4}{Osservatorio Astronomico di Brera, Via Bianchi, 46, I-23807
Merate, Italy}

\altaffiltext{5}{Istituto di Fisica Cosmica ed Applicazioni dell'Informatica,
CNR,
Via U. La Malfa 153, I-90146 Palermo, Italy}

\altaffiltext{6}{Universit\`a dell'Insubria, Como, Via Lucini 3, I-22100
Como, Italy}

\altaffiltext{7}{Space Research Organization of Netherlands, Sorbonnelaan 2,
3584 CA Utrecht, NL, and Astronomical Institute, Utrecht University, P.O.
Box 80000, 3504 TA Utrecht, NL}

\altaffiltext{8}{Astrophysics Division, Space Science Department of ESA,
2200 AG Noordwijk, The Netherlands}

\altaffiltext{9}{Center for Space Research and Department of Physics,
MIT, 77 Mass.\ Ave.,  Cambridge, MA 02139, USA}

\altaffiltext{10}{Osservatorio Astronomico di Roma, Via Frascati, 33, I-00040
Monteporzio Catone, Italy}

%AUTHOR LIST AND ORDER ARE PROVISIONAL

\begin{abstract}
We report on  results of a target of opportunity observation of the X-ray
transient XTE J1118+480 performed on 2000 April 14--15 with the Narrow Field
Instruments (0.1--200 keV) of the \sax\/ satellite. The measured spectrum is a
power law with a photon index of $\sim 1.7$ modified by an ultrasoft X-ray
excess and a high-energy cutoff above $\sim 100$ keV. The soft excess is
consistent with a blackbody with temperature of $\sim 40$~eV and a low flux,
while the cut-off power law is well fitted by thermal Comptonization in a
plasma with an electron temperature of $\sim 10^2$ keV and an optical depth
of order of unity. Consistent with the weakness of the blackbody, Compton
reflection is weak. Though the data are consistent with various geometries of the hot 
and cold 
phases of the accreting gas, we conclude that a hot accretion disk is the most 
plausible model. 
The Eddington ratio implied by recent estimates of the mass and the distance
is $\sim 10^{-3}$, which may indicate that advection is probably not the dominant 
cooling mechanism. 
We finally suggest that the reflecting medium has a
low metallicity, consistent with location of the system in the halo.
\end{abstract}

\keywords{accretion, accretion disks --- binaries: general ---  black hole
physics --- stars: individual (XTE J1118+480) --- X-rays: observations ---
X-rays: stars}

\section {Introduction}
\label{intro}

\begin{figure}[t!]
\epsscale{1.0}
\plotone{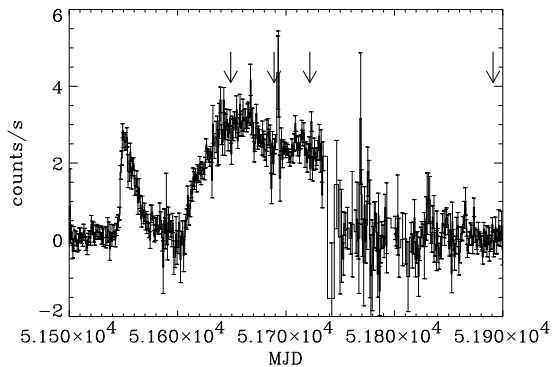}
\caption{One-day-average count rates from \source\/ detected
with the \asm\/ during the outburst. (From the Internet
public archive at xte.mit.edu/XTE/asmlc/ASM.html.)
The \sax\/ observations are marked with the arrows. The observation analyzed
here is the earliest one.
}
\label{f:outburst}
\end{figure}

The high Galactic latitude source ($l = 157.7\degr$, $b = 62.3\degr$) \source\
was discovered in the 2--10 keV energy band with the All-Sky Monitor
(ASM) aboard the {\it Rossi X-Ray Time Explorer\/} (\xte) satellite on 2000
March 29 as a weak ($\sim 40$ mCrab) X-ray source. Retrospective analysis showed that
its flux was slowly rising since 2000 March 5 \cite{Remillard00} and that
another outburst had occurred in 2000 January 2--29, with a similar peak flux on 
January 6 \cite{Remillard00}, see Figure \ref{f:outburst}. The source was also detected
in the 20--100~keV band with the {\it CGRO}/BATSE experiment (Wilson \&
McCollough 2000): the flux was $\sim 60$ mCrab on March 26 and $\sim 110$ mCrab
on January 11. The energy spectra derived from the ASM and BATSE data were
described by a power law with photon index of $\sim1.8$ and $\sim2.1$,
respectively. These power law slopes are typical of black hole binaries (e.g.,
Cyg X-1) in the hard state.

A radio counterpart of \source\ was discovered by Pooley \& Waldram (2000) on
March 31 with a flux density of 6.2~mJy at a frequency of 15~GHz. Also, the
source was detected  in the optical and near-infrared bands
\cite{Garcia00,Chaty00,Wren00} with magnitudes of 12.12, 11.75 and 11.06 in the
J, H, K bands, respectively, on 2000 April 4.2 (UT), and of 13.11 in the V band
on March 28.206. Optical spectroscopy performed on March 31.0 shows H$\alpha$
emission with an FWHM (2200 km s$^{-1}$) that is approximately equal to the
largest found in black hole X--ray novae during quiescence \cite{Garcia00}.
Observations performed with the {\it Hubble Space Telescope\/} (Haswell, Hynes,
\& King 2000a) on 2000 April 8 show an H$\alpha$ absorption line with full
width corresponding to radial velocities likely $> 10^4$~km s$^{-1}$, which are
suggestive of a massive accretor.

The optical counterpart of \source\ is found at R.A.\ (2000) $= 11^{\rm
h}18^{\rm m}10^{\rm s}.85$ and Decl.\ (2000) $=+48\degr 02'12''.9$ (Uemura et
al.\ 2000). A star with a 18.8 $R$ magnitude in the USNO A1.0 and A2.0
catalogues is  within $2''$ of that direction, and it has been proposed as the
quiescent optical counterpart of the source (Uemura et al.\ 2000). Its low
magnitude is consistent with the LMXB nature of the transient source
\cite{Tanaka96}. An interesting property of the system in outburst is the very
low X-ray-to-optical flux ratio ($\sim 5$) with respect to the typical value of
$\sim 500$ found in LMXBs \cite{Jvp95}.

Cook et al.\ (2000) reported the discovery of a sinusoidal periodic variation
of the V-band light curve with a period of 0.1706 days, while a value
of $0.17078\pm 0.00004$~days has been obtained by Uemura et al.\ (2000).
No evidence of periodicity was found in the soft X-rays \cite{Hynes00}.
The above optical modulations have been interpreted as a superhump
\cite{Kuulkers01}, while the true binary period ($P = 0.1703$~days) has been
determined by McClintock et al. (2000, 2001)\nocite{mc00,mc01a} and
Wagner et al. (2000)\nocite{wg00} through radial velocity studies of the
optical counterpart in quiescence.
Revnivtsev, Sunyaev, \& Borozdin (2000) report the discovery of a QPO feature
at 0.085 Hz in the X-ray light curves from the Proportional Counter
Array (PCA) aboard \xte, while the evolution of this QPO for a long time has
been investigated by Wood et al. (2000)\nocite{wd00}.
Besides the QPO, the Power Spectral Density (PSD) shows a strong band--limited
noise component (40\% fractional rms), also typical of BHCs in their low
state.
%A continuum red-noise component in the power spectrum was also
%found in the $\sim 0.03$--70 Hz (rms amplitude of 40\%).
No significant
variability was detected at higher frequencies. Similar variations and QPOs
were observed in the optical/UV band \cite{Haswell00b}.

The definite confirmation of the black hole nature of the source has been
provided by measurements of the mass function being as large as $\sim 6\msun$
(McClintock et al.\ 2000, 2001a, hereafter M01a; Wagner et al.\ 2000),
rivaled
only by that of V404 Cyg. In \S \ref{pars} below, we discuss the black hole
mass and inclination implied by those and other measurements.

Given the broad energy band of the instrumentation on board \sax\ (0.1--200
keV) and the high Galactic latitude of the source, \source\ could be searched
for an ultrasoft component during a hard spectral state with unprecedented
sensitivity. As soon as the source became observable with \sax\/, we started a
Target Of Opportunity (TOO) observational program  devoted to study the 
spectral properties of the source. Here we present results of the first 
observation performed on 2000 April 14--15. In figure 1 we show the epochs
of the source observations with \sax.

\section{Parameters of the System}

\subsection{Masses, inclination and distance}
\label{pars}

McClintock et al.\ (2000), M01a and Wagner et al.\ (2000) have recently detected
absorption lines of an M0--M1 V star in their red optical spectra obtained when
the object was in a quiescent state. The weighted average of their measurements
of the velocity amplitude for the secondary is $K_2 =697 \pm 9$ km s$^{-1}$,
which yields the mass function of
\begin{equation}\label{mass_function}
f(M_2) = \frac{(M_{\rm X}\sin i)^3}{(M_{\rm X}+M_2)^2}=6.0 \pm 0.2\, \msun
\end{equation}
for $P=0.17$ days (where $i$ is the inclination).

At the same time, Garnavich (2000; see also M01a) found a double sinusoidal
variation with a peak-to-peak amplitude of 0.15 mag in the $I$-band light curve
phased to the 0.17-day period. This suggests that the secondary is tidally
distorted.

In systems with mass ratios of $q=M_{\rm X}/M_2 \ga 1.25$, the average density
of a Roche lobe-filling secondary is determined solely by the orbital period,
which implies simple period--mass--radius--spectral type relations for lower
main sequence (MS) secondaries (Frank, King, \& Raine 1992). In the case of XTE
1118+480, the orbital period  implies a spectral type of M2 and a mass of $\sim
0.45 \, \msun$ for a MS secondary with near-solar metallicity, whereas a star
with lower metallicity would have somewhat earlier spectral type (e.g.\
Beuermann et al.\ 1998).  An evolved secondary would have a spectral type
significantly later than that of a zero-age MS star with the same density
(i.e., in a binary with the same orbital period) and a lower mass (e.g.\
Baraffe \& Kolb 2000). If the secondary is a MS star with $M_2 \simeq 0.45\,
\msun$, equation (\ref{mass_function}) yields $q \ga 15$, and $M_{\rm X} \sin^3
i \ga 6.8\, \msun$. The lack of dips or eclipses for a Roche-lobe filling
secondary, in particular in the X-ray light curve (Revnivtsev et al.\ 2000),
yields an upper limit for the inclination of $i < 80\degr$, and $M_{\rm X} \ga
7.1\, \msun$.

The light curve presented in fig.\ 1 of M01a shows some deviations from a purely
ellipsoidal light curve. First, it shows some extra light at $\sim 0.05 P$
after the inferior conjunction of the secondary, which, according to M01a can be
explained by a bright spot at the edge of accretion disk. There is also a phase
lag of 0.05 relative to the spectroscopic ephemeris, which makes the
ellipsoidal nature of the light curve somewhat problematic. On the other hand,
M01a found a significant contamination of the visual light by an accretion disk,
which likely contaminates also the infrared light. In this case, eclipse
effects in an extended disk can possibly shift the phase of the light curve.

Thus, we follow M01a and assume that the light curve is dominated by ellipsoidal
effects. The amplitude of the double sinusoid is then positively correlated
with $q$, $i$, and the Roche-lobe filling factor. Moreover, the two minima of
the ellipticity effect are in general unequal, the one corresponding to the
superior conjuntion of the distorted star being deeper than the other. The
magnitude difference between the two minima is a function of $i$ and the
filling factor. In particular, for a Roche-lobe filling star, this difference
should be about 0.1 mag for $i \geq 80\degr$ decreasing almost linearly to
0 at $i \sim 40\degr$.

The light curve presented by M01a shows the shallower minimum at the inferior
conjunction (as expected) of $\Delta I \sim 0.11$, and the magnitude difference
between the two minima of $\sim 0.07$, indicating $i \geq 60\degr$. On the
other hand, for $q \ga 15$ and a Roche-lobe filling star with convective
envelope, we estimate the amplitude of the shallower minimum to be $\Delta I
\ga 0.35 \sin^2 i $ mag, which suggests much lower inclination, $i \simeq
34\degr$. This shows that the amplitude of the ellipsoidal variability is
reduced due to contamination of the $I$-band flux by some additional
contribution from other source(s), e.g.\ an accretion disk.

M01a found that the secondary contributes $42\%$ (K5V) to $26\%$ (M1V) of the
total flux at 5900 \AA. To evaluate a contribution of the accretion disk at the
$I$-band, we assume that the local disk spectrum corresponds to either
$F_\nu$ or $F_\lambda$ being constant. Assuming an M1--M2 secondary, we
estimate the contribution of the disk as $\sim 30\%$ and $\sim 60\%$, in the
first and second case, respectively.

Then, the following equations can be used to determine $q$, $i$, and the binary
separation, $a$. Kepler's third law reads,
\begin{equation}
M_{\rm X}+M_2=0.01343 {(a/{\rm R}_{\sun})^3\over  (P/1\,{\rm day})^2} \msun.
\end{equation}
The secondary's radial velocity amplitude, $K_2 = 697$ km/s, corresponds to
\begin{equation}
a \sin i = 2.34 \frac{1+q}{q} {\rm R}_{\sun}.
\end{equation}
The full (peak-to-peak) amplitude produced by the tidal distortion is
\begin{equation}
\Delta I = 4.3 q \left(R_{\rm t}\over a\right)^3 \sin^2 i\, {\rm mag},
\label{eq:Delta}
\end{equation}
where $R_{\rm t}$ corresponds to the ``transverse radius''  of the secondary,
which is about 96\% of the mean radius, $R_2$, of its Roche lobe. The numerical
coefficient in equation (\ref{eq:Delta}) was evaluated from synthetic $I$-band
light curves generated with the Wilson-Devinney code (Wilson 1990) using a
gravity-darkening exponent of 0.32, a linear limb-darkening coefficient of 0.6,
and using the shallower minimum in the synthetic light curve. Assuming a MS
secondary with $M_2 \simeq  0.45 \msun$ and $R_2 \simeq 0.45  {\rm R}_{\sun}$,
we obtain $a=3.5 {\rm R}_{\sun}$, $q=44$, $i=43\degr$, $M_{\rm X}=20\msun$,
and $a=2.8{\rm R}_{\sun}$, $q=22$, $i=61\degr$, $M_{\rm X}=10\msun$ in the
cases with constant $F_\nu$ and $F_\lambda$, respectively. The first solution
produces equal minima, which is not observed, whereas the second one results in
a too shallow second minimum.

On the other hand, the presence of extra light near the first minimum will
reduce its depth, whereas the second minimum may be not affected. Using only
the depth of the second minimum, we get $a=3.13 {\rm R}_{\sun}$, $q=31$,
$i=50\degr$, $M_{\rm X}=14\msun$ in the first case, and a solution similar to
that of M01a, $a=2.53{\rm R}_{\sun}$, $q = 16$, $i = 80\degr$, $M_{\rm
X}=7.1\msun$,  in the second case. Note that the first solution still 
produces equal minima in the light curve.

We note that the determinations of the binary parameters by M01a and above
supersede analogous constraints of Dubus et al.\ (2001), obtained before the
recent measurements of $K_2$. Those authors used instead the peak-to-peak
separation of very broad double-peaked H{\sc i} Balmer and He{\sc ii} 4686 \AA\
emission lines, measured by them as $\Delta v \approx 1270$, 1470 km/s,
respectively. They then assumed that $\Delta v$ represents the projected
Keplerian velocity of the outer disk, and used a theoretical value of its
radius, $R_{\rm d}$. For a Keplerian disk with a power-law emissivity, the
relation,
\begin{equation}
1+q^{-1}=\left(\frac{\Delta v}{2 K_2}\right)^2 {R_{\rm d}\over a},
\end{equation}
is satisfied (e.g.\ Smak 1981). However, with the measured $K_2$,  $1+q^{-1}
\approx 0.8 R_{\rm d}/a$, $1.1 R_{\rm d}/a$ for the Balmer and He{\sc ii}
lines, respectively. This implies $R_{\rm d} \sim a$ for any value of $q$,
which then appears to invalidate their adopted assumption of a Keplerian outer
disk.

Summarizing, the present most likely range of the mass and inclination is
\begin{equation}\label{im}
60\degr\la i \la 80\degr, \quad 10\la M_{\rm X}/\msun \la 7.
\end{equation} 
The lower and upper limit on $i$ follow from the unequal light-curve minima and the lack of 
eclipses, respectively.

The presence of the M1--2 V star with $V \simeq 20.4$ (M01a) implies the
distance of
\begin{equation}
d\simeq (1.3\! -\! 1.9)\, {\rm kpc},
\end{equation}
where the error also takes into account the uncertainty  in the interstellar
absorption (see below). We also assumed $M_V \simeq 8.96$--9.9 for such a MS
star with near-solar metallicity (Baraffe \& Chabrier 1996). Note that $d$ will become 
lower for lower metallicity (Beuermann et al.\ 1998).

\subsection{The hydrogen column density}
\label{nh}

An accurate determination of the hydrogen column density, $\nh$, is of key
importance for determining the correct spectral model for the source emission.
Given the high Galactic latitude of the source, the total Galactic column
density along the source direction is likely to be close to $\nh$, as well as
it represents an upper limit to it. The Galactic column density of H{\sc I} has
been measured in radio in 5 directions within $1\degr$ of the direction of the
optical counterpart, yielding the range of $N_{\rm H,G}=(1.28$--$1.44) \times
10^{20}$ cm$^{-2}$ (Dickey \& Lockman 1990, as implemented in the {\sc ftools}
software package). The measurement in the direction closest ($0.29\degr$) to
that of \source\ gives $1.28 \times 10^{20}$ cm$^{-2}$, and the average
weighted by the angular distance from the source direction is $1.34 \times
10^{20}$ cm$^{-2}$. We note that, at these low values of $\nh$, absorption is
effective only at soft X-rays, at which H{\sc i} is indeed the main absorbing
species. 
A lower Galactic column density ($0.75\times 10^{20}$ cm$^{-2}$) along the
line of sight is derived  following Schlegel, Finkbeiner \& Davis (1998),
who obtained full--sky dust maps on the basis of the {\it COBE} and {\it IRAS}
satellite data. The  uncertainty in their reddening maps is estimated 
to be about 16\%. Taking into account that the $\nh$ value along the line 
of sight, obtained by interpolation from these and  H{\sc i} maps, can differ
from the real value by a factor $\sim 2$ (Faison et al. 1998), we can state
that the above $\nh$ estimates are consistent each other.

Then, Dubus et al.\ (2001) have found  3 weak absorption features 
(Ca{\sc ii} 3933 \AA\/ doublet and 3968 \AA), interpreted as coming 
from 3 interstellar clouds in the
line of sight moving at different velocities. From best-fitting, they obtain
the total $N_{\rm CaII} \simeq 1.4\times 10^{12}$ cm$^{-2}$ (without giving its
uncertainty). They then convert this value to $\nh$ using an average range of
$\langle \log (\nh/N_{\rm CaII})\rangle \approx 8.1$--8.5 obtained for
high-altitude lines of sight (Sembach \& Danks 1994). This yields $\nh$ in the
range of (1.76--$4.4)\times 10^{20}$ cm$^{-2}$. We note, however, that it is
sufficient that the actual $\nh/N_{\rm CaII}$ ratio in the direction of
\source\ is just $\sim 25\%$ less than the lower limit of the above {\it
average\/} ratio range to achieve the consistency of $\nh$ from radio mapping
of H{\sc i} with that from Ca{\sc ii}. In fact, the H and K lines of Ca{\sc ii}
correlate poorly with $\nh$ in general and are thus considered not suitable for
the purpose of accurate determination of $\nh$ (Dickey \& Lockman 1990 and
references therein).

On the other hand, Hynes et al.\ (2000) obtained $\nh$ in the range of
(0.35--$1.15) \times 10^{20}$ cm$^{-2}$ by fitting their 0.1--0.2 keV {\it
EUVE\/} spectrum by a power law with a range of indices. We note that such
narrow-band fits appear to be easily subject to systematic errors due to the
possibility that the actual spectrum is not a power law. Our \sax\/ data
include this range, and we have found $\nh \sim (0.7$--$1.5) \times 10^{20}$
cm$^{-2}$ by fitting a range of models. Power-law models gave lower values of
$\nh$, consistent with Hynes et al.\ (2000), but models with blackbody and
Comptonized blackbody spectra gave higher $\nh$, consistent with the
measurements of Dickey \& Lockman (1990). Given the results above, we consider
the most likely range of $\nh$ to be $\sim 1$--$1.5 \times 10^{20}$ cm$^{-2}$.

\section {The Observation and Spectral Data}
\label{obs}

The observation was performed with the \sax\ Narrow Field Instruments (NFIs).
The \asm\/ light curve of \source\/ in the 2--10 keV energy band with the mark
of the \sax\ observations is shown in Figure~1. The NFIs embody a Low Energy
Concentrator Spectrometer (LECS, 0.1--10 keV, Parmar et al.\
1997\nocite{Parmar97}), 2  Medium Energy Concentrators Spectrometers (MECS,
1.3--10 keV, Boella et al.\ 1997), a High Pressure Gas Scintillator
Proportional Counter (HPGSPC, 3--100 keV, Manzo et al.\ 1997\nocite{Manzo97}),
and a Phoswich Detection System (PDS, 15--300 keV, Frontera et al.\
1997\nocite{Frontera97}). The LECS and MECS have imaging capabilities, while
the HPGSPC and PDS are collimated detectors with a Field of View (FOV) of
$1^{\circ}$ and $1.3^{\circ}$, respectively. The PDS instrument makes use of
rocking collimators for background monitoring.

We observed the source from 2000 April 14.489~UT to 15.528~UT. The source
exposure times were
21197~s, 29758~s, 42138~s, and 20146~s for the LECS, MECS, HPGSPC and PDS,
respectively. Useful data were selected from time intervals that met the
following criteria: the satellite outside the South Atlantic Geomagnetic
Anomaly, the elevation angle above the Earth limb by $\geq 5\degr$, dark
Earth (for LECS), stabilized high voltage supplies.

The source was detected with all the NFIs, from $\sim$0.1 to 200~keV. The LECS
and MECS source spectra were extracted from a region with a radius of $8'$
around the centroid of the source image. As background spectra we used standard
files obtained from the observation of blank fields. The 2 MECS spectra were
equalized and co-added. We used background spectra accumulated from dark Earth
data for the HPGSPC (note that the instrument's collimator was kept on the
source for the entire observation). The background level of the PDS was
estimated by swapping its collimators off source every 96 s. The energy bands
used for spectral fitting were limited to those where the response functions
were best known, i.e, 0.12--4 keV, 1.7--10 keV, 7--29 keV, and 15--200 keV, for
the LECS, MECS, HPGSPC and PDS, respectively. 
The count rate spectra  were analyzed using the {\sc xspec} software
package \cite{Arnaud96}. A systematic error of 1\% was added in quadrature
to the
statistical uncertainties of the spectral data, on the basis of the calibration
results obtained with the Crab Nebula, that was observed 4 days before
our observation (10 April 2000). We allowed for free normalization of the 
instruments in
multi-instrument fits with respect to MECS. For clarity of display,
the unfolded spectra from multi-instrument fits were renormalized to the
MECS level.

The results below have been obtained with the spectra averaged over the entire
observation. We have also performed a spectral analysis on different time
intervals, but have not found any statistically significant variation of
the  best-fit parameters with time.

The quoted errors for the spectral parameters correspond to 90\% confidence for
one parameter ($\Delta \chi^2 = 2.71$). The elemental abundances are with
respect to those of Anders \& Ebihara (1982), and the opacities are from
Morrison \& McCammon (1983). In calculations of Compton reflection and in
Comptonization models in planar geometry, we assume an inclination of
$i=70\degr$ (see \S \ref{pars}).

\section{Results}
\label{results}

%
% Figure 2
%
\begin{figure}[t!]
\epsscale{1.0}
\psfig{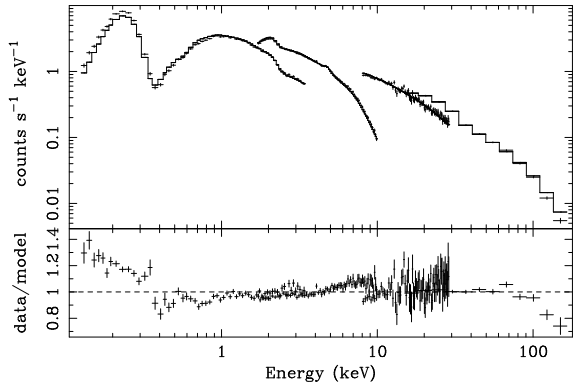}
\caption{{\it Top panel}: The 0.1--200~keV count rate spectrum of \source.
{\it Bottom panel}: Ratio between the count rate spectrum and the power--law  model
absorbed by the Galactic $\nh$. A soft X-ray excess and a high-energy cutoff are 
clearly highly significant.}
\label{f:pl}
\end{figure}

\begin{figure}[!t]
\epsscale{1.0}
\plotone{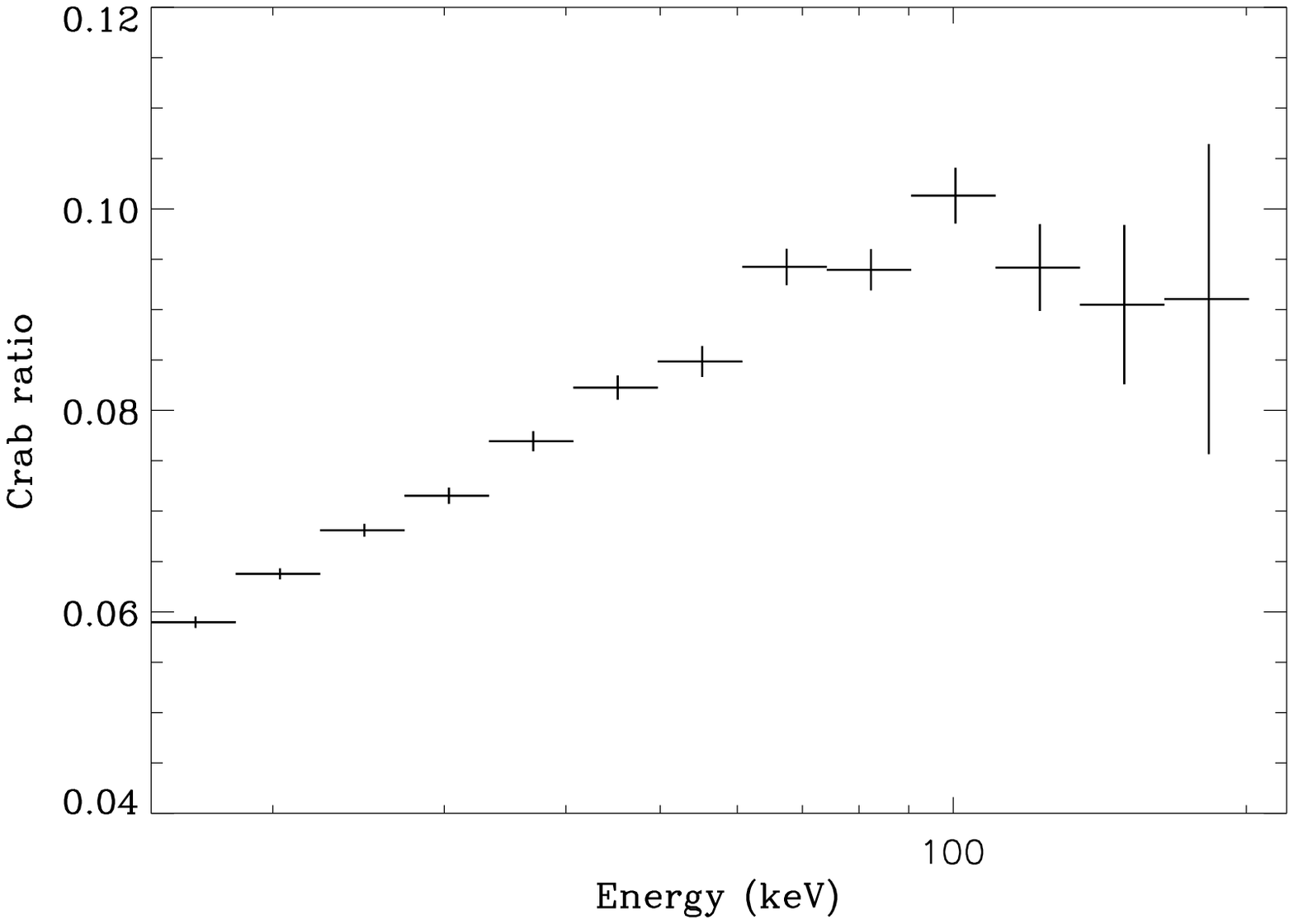}
\caption{Ratio between the count spectrum of \source\ and that of Crab
observed with \sax\/ 4 days before our TOO measurement. The break in the
ratio at $\sim 100$ keV is apparent.}
\label{f:crab_ratio}
\end{figure}

\begin{figure}[t!]
\epsscale{1.0}
\psfig{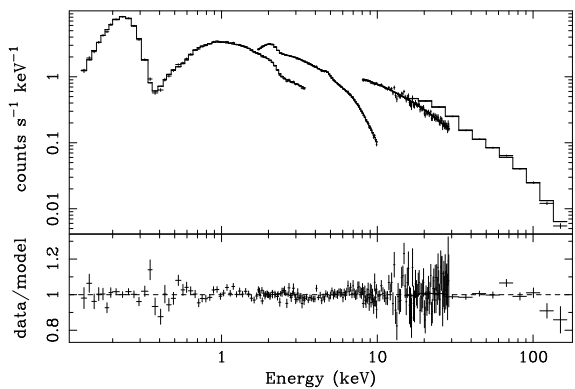}
\caption{{\it Top panel}: The 0.1--200~keV count rate spectrum of \source,
superposed with the best fit model consisting of an absorbed {\sc bb} plus
{\sc cutoffpl} (see text).
{\it Bottom panel}: Ratio between the count spectrum and the model.}
\label{f:cutoffpl}
\end{figure}

The 0.1--200~keV count rate spectrum obtained with the NFIs  (see Figure 
\ref{f:pl}, top panel) was fit with different models. A simple
power-law model ({\sc pl}) with $\nh= 1.28 \times 10^{20}$ cm$^{-2}$
(see \S \ref{nh}) gives an
unacceptable fit with the photon index, $\Gamma= 1.81$, and a very large 
$\chi^2/\nu = 1143/231$, caused mostly by strong systematic residuals at the 
lowest and highest energies, corresponding to a soft X-ray excess at $\la 0.5$ 
keV and a high-energy cutoff at $\ga 100$ keV, respectively, as shown in the 
bottom panel of Figure \ref{f:pl}. The high energy cutoff is also apparent 
from figure \ref{f:crab_ratio}, where we show the ratio of the \source\/ 
spectrum measured by the PDS  with the corresponding one of the Crab Nebula, 
which was observed for cross-calibration purposes 4 days before our 
observation. Replacing the power law with an e-folded power law ({\sc cutoffpl}) 
and maintaining the $\nh$ value fixed to $1.28 \times 10^{20}$ cm$^{-2}$ 
only marginally improve the fit ($\chi^2/\nu = 1136/230$): 
the systematic residuals at low and high
energies are still there (see Figure \ref{f:pl}), and the best fit cut--off 
energy is very high and unconstrained. 
Only allowing a free $\nh$ the fit does improve
significantly, to $\chi^2/\nu = 319/229$, but only for
an $\nh = 0.66^{+0.03}_{-0.03} \times 10^{20}\,$cm$^{-2}$, about
a half of $N_{\rm H,G}$, even if consistent with the $\nh$ based on the
dust IR maps (\S \ref{nh}).
However still there remain strong systematic residuals at the lowest and 
highest energies, as confirmed by the still unacceptable $\chi^2$.
Thus we added to the previous model a blackbody spectrum ({\sc bb}).
With this additional component the fit quality significantly increased
($\chi^2/\nu = 272/227$), with a chance probability that the lower $\chi^2$
is due to chance of $1.4 \times 10^{-8}$. Even if the 
{\sc cutoffpl} model component is not the best description of the high 
energy cut--off (see below), it represents a great improvement with respect 
to a {\sc pl} component, that gives a much worse and unacceptable
quality of the fit ($\chi^2/\nu = 388/228$). This result confirms the
statistical significance of the high energy cut--off in our data.
The best fit model with {\sc bb} plus {\sc cutoffpl} provides an $\nh$
close to $N_{\rm H,G}$ with a rather low
temperature ($kT_{\rm bb}\simeq 50$ eV) of the {\sc bb} spectrum, see Table 1.
Figure \ref{f:cutoffpl} shows the best fit curve along with the residuals of 
the data to the model.

%
% Figure 3
%

%
% Table 1
%
\begin{table*}
%\epsscale{1.0}
\vspace {-2cm}
\psfig{file=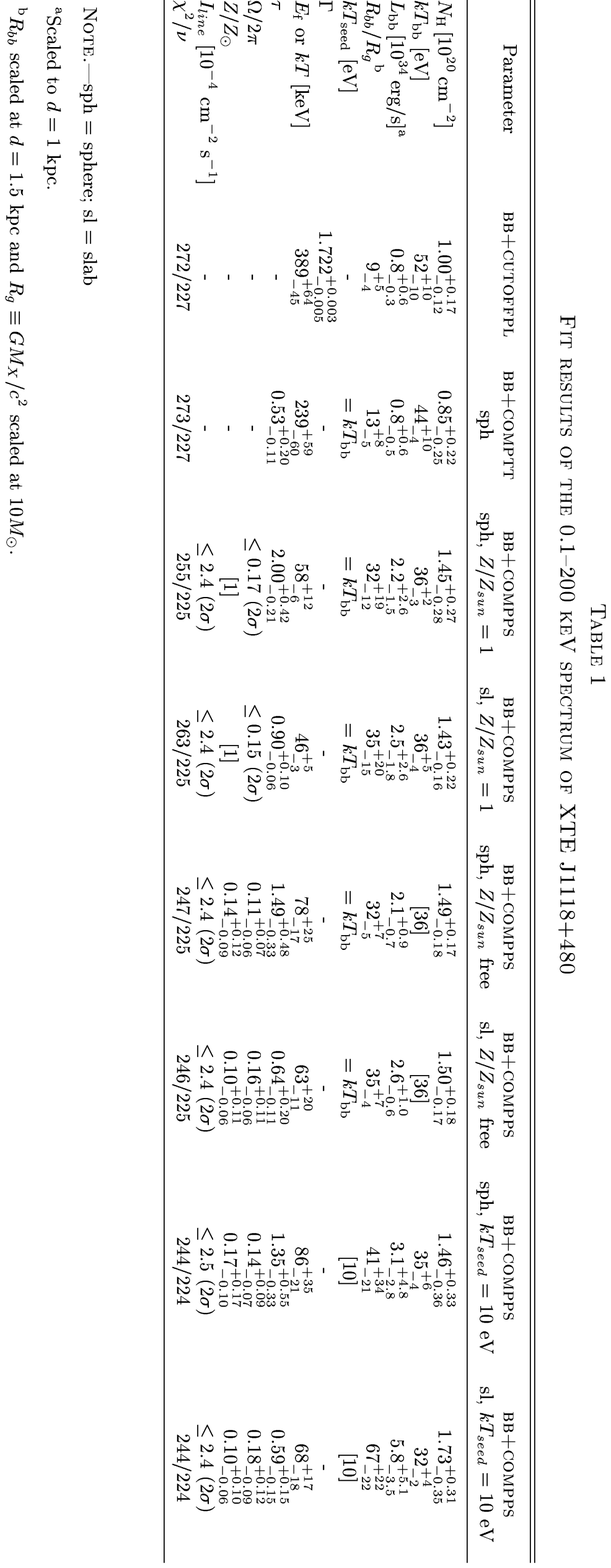,angle=0}
%\plotone{table_1.ps}
\label{t:table1}
\end{table*}

%
% Figure 4
%

\begin{figure}[t!]
\epsscale{1.0}
\psfig{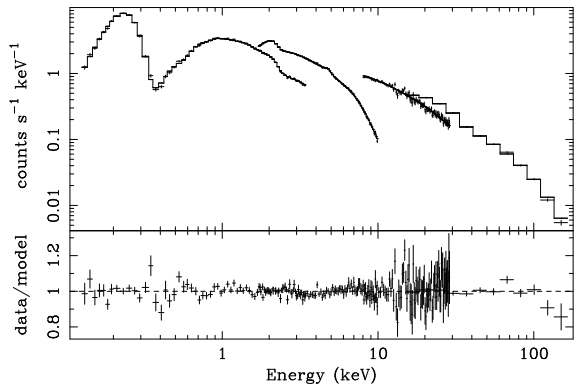}
\caption{{\it Top panel}: The 0.1--200~keV count rate spectrum of \source,
superposed with the best fit model consisting of an absorbed {\sc bb} plus
{\sc comptt} in the spherical geometry (see text).
{\it Bottom panel}: Ratio between the count spectrum and the model.}
\label{f:comptt}
\end{figure}

Still, even the last fit shows strong systematic residuals at the highest 
energies, as an e-folded power law does not reproduce the actual shape of the 
high-energy cutoff in the data. Therefore, we have tested models with the 
main continuum component given by Comptonization, a common physical process in 
compact sources (see Zdziarski 2000 for a review). We first consider the 
thermal-Compton model ({\sc comptt}) of Titarchuk (1994), in which a Wien 
spectrum undergoes scattering in either spherical or slab geometry, and 
relativistic effects are taken into account. We have found that fits with very 
similar electron temperature, $kT$, and $\chi^2$ are obtained for either 
geometry. Similarly to the case of an e-folded power law, the absorbed
{\sc comptt} model yields an implausibly low $N_H$ 
($0.51^{+0.08}_{-0.18} \times 10^{20}$ cm$^{-2}$) with a 
$\chi^2/\nu = 283/228$, similar to that found with the last
model ({\sc bb} plus {\sc cutoffpl}). Adding an absorbed {\sc bb} component,
with $T_{\rm bb}$ equal to the temperature of the seed photons,
the best fit $\nh$ is closer to $N_{\rm H,G}$, even though the fit
quality does not improve ($\chi^2/\nu = 279/227$ or 280/227 according to
the geometry of the electron cloud, a sphere or a slab, respectively; see
Table 1 for the sphere). The cause of it is  that the derived electron
temperature, $kT = 239^{+59}_{-60}$ keV for a sphere and
$kT = 251^{+86}_{-36}$~keV for a slab, is too much large to account for the
observed cutoff at an energy of $\sim 100$ keV (see also Figure 
\ref{f:comptt}).

%
% Figure 5
%

\begin{figure}[t!]
\epsscale{1.0}
\psfig{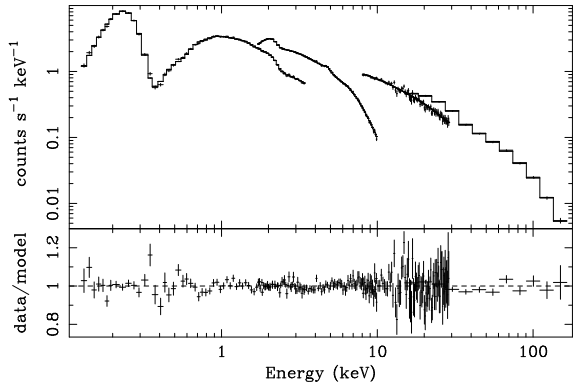}
\caption{{\it Top panel}: The 0.1--200~keV count rate spectrum of \source,
superposed with the best fit model consisting of an absorbed {\sc bb} plus
{\sc compps} model in the spherical geometry and a solar metal abundance 
(see text).
{\it Bottom panel}: Ratio between the count spectrum and the model.}
\label{f:compps}
\end{figure}

Then we tested the
thermal-Compton model ({\sc compps} v3.4\footnote{{\sc compps} is available
on the internet at ftp://ftp.astro.su.se/ pub/juri/XSPEC/COMPPS.}) of Poutanen \& 
Svensson (1996). The advantage of this model is that, in addition to allow for
different geometries of the electron cloud, it allows for the
presence of Compton reflection (Magdziarz \& Zdziarski 1995) from a cold
medium (like an accretion disk) with variable element abundances
with respect to the solar ones. 
We considered spherical or slab geometries with blackbody seed
photons distributed homogeneously in the plasma and the Thomson optical depth, 
$\tau$, corresponding to the radius or the half-thickness, respectively.
Initially we assumed reflection from a medium with solar abundances.
For consistency, we also allowed for a narrow Gaussian line ($\sigma =
0.01$ keV) at 6.4 keV. In the spherical geometry and allowing the presence
of additional blackbody photons (apart from
those irradiating the plasma), we find $\chi^2/\nu =255/225$ at 
$kT=58^{+12}_{-6}$ keV, see Table 1. Thus the best fit provides a lower
value of the electron temperature, that allows a better description
of the observed high energy cutoff with respect to the {\sc comptt}
model\footnote{Note that the values of $kT$ obtained with {\sc compps}
are much lower from that derived from {\sc comptt}, which is caused by 
inaccuracies in the shape of the high-energy cutoff in the latter model, see, 
e.g., Zdziarski, Johnson, \& Magdziarz (1996).}, see Figure \ref{f:compps}.
An F-test for 2 additional parameters provides a value of $4.6\times 10^{-4}$ 
for the probability that the $\chi^2$ reduction from {\sc bb + comptt}  
to {\sc bb} + {\sc compps} is due to chance.
 
Similarly we cannot exclude the slab geometry for the same model, that
provides a statistically similar fit, $\chi^2/\nu=263/225$, even if the
latter model seems to show in correspondence of the 150--200 keV data point
a positive residual that is not observed in the case of
the spherical geometry (see Figure~\ref{f:compps}).

%
% Figure 6
%

No evidence of relative reflection  was found
with the above model in both geometries, with a 2$\sigma$ upper limit of
$\sim$0.16 (see Table 1). This result, caused by the absence of an Fe K edge,
contrasts other black black hole binaries in the hard state (e.g.,
Gierli\'nski et al.\ 1997). To check the robustness of the absence of
reflection, we first consider the dependence on the inclination.
We find no evidence of relative reflection even at $i=79\degr$
(about the maximum allowed by the data, \S \ref{pars}) with
a $2\sigma$ upper limit of 0.26. Also no evidence of
a narrow Fe K line is found in the data, with a $2\sigma$ upper limit of
the line intensity of $2.4\times 10^{-4}$ photons~cm$^{-2}$~s$^{-1}$  with
energy from 6.4 to 6.9 keV, independently of the geometry adopted
(see Table~1).

\begin{figure}[!t]
\epsscale{1.0}
\plotone{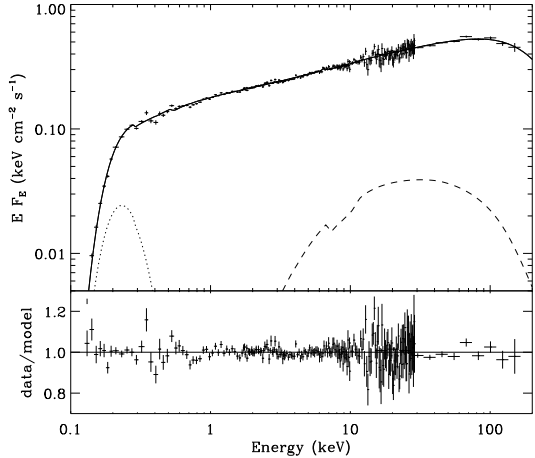}
\vspace{0.5cm}
\caption{The \sax\/ spectrum (crosses) fitted by (solid curve) a blackbody
at 36~eV plus a thermal
Comptonization in the slab geometry and a Compton reflection from a
low-metallicity medium. The latter is shown separately by the dashed curve
while the blackbody is shown by a dotted line. The spectrum is
normalized to the level of the MECS (see text). The bottom panel shows the fit
residuals.
} \label{f:EFE} \end{figure}

A low metal abundance, $Z$, of the reflecting medium is expected for halo
stars, like XTE J1118+480. Spectrally, it has an effect of reducing the K
edge in the reflected spectral component without changing the shape of its
high-energy cutoff. We find that allowing for $Z<Z_\sun$ only slightly
improves the fit  bringing $\chi^2/\nu$ to 247/225 (246/225) in the
spherical (slab) geometry.
Table 1 shows the best fit parameters in both geometrical configurations. 
We are not capable to discriminate between these geometries.
We now observe a statistically significant Compton reflection, similar to the
upper limits found when $Z/Z_\sun$ was fixed to 1.
We notice  that the Comptonization parameters are slightly lower than those 
found with Cyg X-1 in the hard state 
(Gierli\'nski et al.\ 1997), in the assumption of a single temperature 
isotropic model. The fit with variable $Z$ is shown in 
Figure \ref{f:EFE}, which also shows the remaining small
residuals. 

We have also considered a multi-component model consisting of the {\sc compps}
model in which the spectrum of seed photons peaks below the observed range
($kT_{\rm seed}$ fixed at 10 eV), plus an additional {\sc bb} component
plus the narrow Gaussian profile (see above) at 6.4~keV. Such a low energy 
of seed photons is expected, e.g., when Comptonization of thermal synchrotron
emission dominates (e.g.\ Wardzi\'nski \& Zdziarski 2000),
see \S \ref{theory}. This model, in both geometries (sphere or slab),
yields a good description of the data as the last model, with a
$\chi^2/\nu=244/224$. The best fit parameters are reported in Table~1.
%
% and $kT=73^{+42}_{-15}$ keV, $L_{\rm bb}=  (7^{+5}_{-4}) (d/1\,{\rm kpc})^2
% \times 10^{34}$ erg s$^{-1}$, $Z/Z_\sun=0.13^{+0.08}_{-0.06}$,
%
Notice the value of $N_{\rm H}$  in the slab geometry 
($= 1.73^{+0.31}_{-0.35} \times 10^{20}$ cm$^{-2}$) which is somewhat 
above $N_{\rm H,G}$.

%
% Figure 7
%

%
% Figure 8
%
\begin{figure}[!t]
\epsscale{1.0}
\plotone{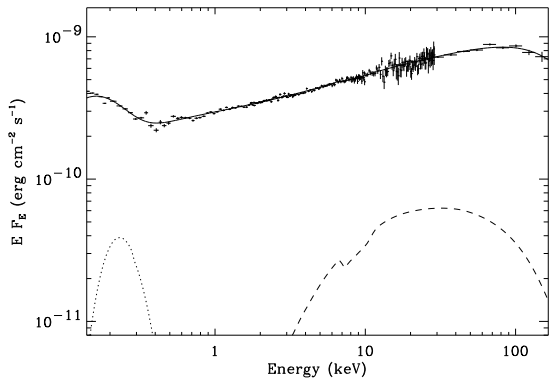}
\caption{The absorption-corrected model spectrum (solid curve)
corresponding to the fit shown in Fig.\ \ref{f:EFE}. The dashed curve shows
the Compton reflection component, while  the dotted curve shows the blackbody
spectrum observed. The initial
(i.e., before Compton scattering) spectrum of the seed photons incident on
the plasma corresponds to  the peak at $\sim$0.15~keV.}
\label{f:model}
\end{figure}

Leaving the metallicity free to vary, we have also tested for the effect of
replacing the blackbody spectrum by a disk blackbody, in this case
assuming the seed photons for Comptonization to be also those of the disk
blackbody. This model yields plasma parameters very similar to
the blackbody case above and a similar fit quality ($\chi^2/\nu= 244/224$
for a slab and $= 245 /224$ for a sphere). The inner disk temperature
is  $kT_{\rm in}=41^{+3}_{-4}$ eV at an inner radius greater than $\sim$500~km
(scaled at $d = 1.5\,{\rm kpc}$), that corresponds to $\sim34 R_g$ ($R_g 
\equiv G M_X/c^2$), assuming $M_X = 10 M_\sun$.
%
%
%The models with low metallicity have rather weak Fe K$\alpha$ lines. The
%model shown in Figure \ref{f:EFE} has a line with the equivalent width of
%$4_{-4}^{+14}$ eV. This range is consistent with corresponding theoretical
%predictions for the fitted reflection and metallicity.

The absorbing columns fitted in models using {\sc compps} are close the
Galactic $\nh$, with some absorption intrinsic to the source possible as well.
We note that the value of $2.8 \times 10^{20}$ cm$^{-2}$ favored by Dubus et
al.\ (2001) is inconsistent with our data, yielding, e.g., $\chi^2/\nu = 
308/226$
in the model with Comptonization in the spherical geometry. This confirms our
independent conclusions of \S \ref{nh}.

From the overall best-fit models, the mean flux level of the source during the
observation is $5.7 \times 10^{-10} \,$\ergcms, $6.9 \times 10^{-10}
\,$\ergcms, $1.9 \times 10^{-9} \,$\ergcms, and $3.4 \times 10^{-9} \,$\ergcms\
in the 0.1--2~keV, 2--10~keV, 15--200~keV, and 0.1--200 keV bands, respectively
(normalized to the MECS detector). The total (0.01--$10^3$ keV) flux  corrected
for absorption is somewhat model-dependent, and it equals to $4.2 \times
10^{-9} \,$\ergcms\ for the {\sc bb + Compps} model in the slab configuration
($Z/Z_{sun}$ free to vary in the fit) (see Table 1 and fig.\ref{f:EFE}). 
The absorption-corrected flux of the observed blackbody is  $2.2 
\times 10^{-10} \,$\ergcms\ to be compared with a flux of $3.8 \times 10^{-10} 
\,$\ergcms\ due to blackbody photons  incident on the hot plasma, i.e. 9\% 
of the total flux (with an apparent amplification factor of 11).

A temporal analysis of the data with Fourier techniques shows that the
fractional variation of the source flux in the 0.01--50~Hz frequency range
and the 2--10~keV energy band is 42\% rms, which is consistent with the
findings by Revnivtsev et al.\ (2000). In the 0.1--2~keV range, the
fractional variation is even higher (62\% rms). Also, a QPO with the
centroid frequency at $\sim 0.08$ Hz is apparent in our power spectral
density estimate. We do not find any statistically significant modulation
at the optical period of 0.17~day. This implies that the X-ray
source is axially symmetric and no absorption in the line of sight is
associated with the ouftlow of matter from the companion star.

\section{Spectral comparison with \xte\ data}
\label{comparison}
\xte\/ performed many short pointings of \source.  McClintock et al.
(2001b) reported spectral results on this source  based on the 
observation of April 18.8111--18.9583 UT and, for the study of the high
energy part of the spectrum ($>$100~keV), results of the summed spectra
obtained between April 13.39 UT and May 15.39~UT. No evidence of a high
energy cut--off was found in the summed spectra, while the high energy data
of the April 18 observation were affected by large uncertainties (see
figure 4 of McClintock et al. 2001).

One of the \xte\/ pointings of \source\/ was performed simultaneously
with a part of our observation. The \xte\/ observation
(see, e.g., Revnivtsev et al.\ 2000) started 2000 April 15 7:57:52 UT, and
had an exposure time of 1152 s for the PCA, and 378 s and 361 s for
the 2 High Energy X--ray Timing Experiment (HEXTE) clusters, respectively.
A  detailed study of this and other \xte\/ observations of \source\ will be
given elsewhere.
We first fitted the \xte\/ spectrum by an e-folded power
law, obtaining $\Gamma\simeq 1.8$, i.e., a higher index
than $\Gamma\simeq 1.7$ for the \sax\/ spectrum (Table 1). 
This difference is likely to be of instrumental  origin, as a similar difference between
the PCA and other X-ray instruments is commonly found in fits to the Crab
as well as other sources (e.g., Gierli\'nski et al.\ 1999; Done,
Madejski, \& \.Zycki 2000).

Then, we fit the 3--200 keV data with Comptonization in the spherical
geometry, Compton  reflection at $Z_\sun$, and a narrow Fe K$\alpha$ line.
We obtain $\Omega/2\pi= 0.05$ at the best fit, confirming the weakness of
Compton reflection.  However, the fitted electron temperature is somewhat
larger than that of the \sax\/ data, $kT= 190^{+50}_{-80}$ keV. We find,
however, that if apply a correction of $\Delta\Gamma =0.1$ to the model
fitted to the PCA data, $kT$ becomes equal to $140^{+60}_{-100}$ keV,
consistent with the values of $kT$ in Table 1. The large confidence regions
of $kT$ are due to the limited statistics of the HEXTE data (the \xte\/ 
observation lasted only a small fraction of the \sax\/ one).
The fact that the summed \xte\/ spectra (covering more than a month elapsed
time) are still described by a power law with no high energy cut--off
below $\sim 150$ keV (McClintock et al. 2001) could be the result
of the evolution of the source spectrum. In fact our 
preliminary results of the later 2 observations of the source (see fig. 1) 
show a decreasing evidence of the high energy cut--off with time
(Frontera et al. 2001b).

Summarizing, also the \xte\/ data appear to be compatible with the \sax\/ ones,
especially when the difference in the calibration between the PCA and 
\sax\/ is taken into account.

\section{Theoretical Implications}
\label{theory}

The Eddington luminosity at an H mass fraction of 0.7 can be written as
\begin{equation}
L_{\rm E}\simeq 1.5\times 10^{39} (M_{\rm X}/ 10\msun)\, {\rm erg\, s}^{-1},
\end{equation}
and the gravitational radius is
\begin{equation}
R_{\rm g}\equiv GM_{\rm X}/c^2 \simeq 15  (M_{\rm X}/ 10\msun)\, {\rm km}.
\end{equation}
The total X-ray luminosity corrected for absorption calculated using the
Comptonization model shown in Fig.\ \ref{f:model} and assuming isotropy is
\begin{eqnarray}
L_{\rm X} \simeq & \!\!\!\!\!\!\! 1.1\times 10^{36} (d/1.5\, {\rm kpc})^2\, {\rm erg\, s}^{-1} \nonumber\\
\simeq &  8\times 10^{-4}\! \left( d\over 1.5\,{\rm kpc}\right)^2\!\! \left(
M_{\rm X} \over 10\msun\right)^{-1}\!\! L_{\rm E}.
\end{eqnarray}
Here we used the estimates for the black hole mass and the source
distance (\S \ref{pars}). We note that the $L_{\rm X}$ above does not represent
the bolometric luminosity of the source, whose $EF_E$ spectrum has been
observed by Hynes et al.\ (2000) to have a maximum in the far UV at a level
similar to that of the peak at $\sim 100$ keV. Given their results, the
bolometric luminosity can be estimated as $\sim 2L_{\rm X}$. Still,
$L_{\rm X}$ represents the luminosity of the hot plasma flow.

A fundamental issue is the geometry of the source. The weakness of Compton
reflection and of blackbody emission poses some contraints. There are 
two possible solutions we can envisage. The direct emission from the 
accretion disk is smeared by Compton scattering in the corona, or the 
hot gas subtends a small solid angle by a cold medium.

A static disk/corona geometry (e.g., Haardt \& Maraschi 1991), 
similary to the case of Cyg X-1 in the hard state, can be reconciled with the
data,
as the large value of $\tau/\cos i$ would scatter most of the reflection and 
blackbody emission in the power law continuum. However, because of the tight 
coupling of the hot and cold phases, this model cannot produce a spectrum
as flat as
that observed. In principle, a mildly relativistic coronal outflow can
alleviate
the problem (Beloborodov 1999; Malzac, Beloborodov, \& Poutanen 2001).

Given that, the specific coronal model of Merloni, Di Matteo, \&
Fabian (2000) has some difficulties. It can produce a flat spectrum, but, 
as it postulates magnetic flares in a disk-corona geometry, it
predicts the presence of a strong blackbody component about an order of
magnitude above that observed, even with relativistic bulk motion of the flares
taken into account (compare their Fig.\ 1 with our Fig.\ \ref{f:model}).

On the other hand, the observed spectrum is fully compatible with Compton
scattering in a hot central disk. Indeed, this is the geometry of our overall best-fit
model. Then, the weakness of the blackbody and Compton--reflection components is
naturally accounted for by a small solid angle subtended by any cold medium as
seen from the hot disk. An important issue here is the origin of the seed
photons. The weakness of the blackbody corresponds to rather small size of the
cold region. In either the blackbody or disk-blackbody model (see \S
\ref{results} and Table 1), the characteristic radius of the blackbody emitter is 
$\ge \sim 30 R_{\rm g}$ (at $d=1.5$ kpc, $M=10\msun$) for our best fit models. 
Thus, the seed photons can be supplied by an outer cold disk, probably overlapping 
with the hot one.

Another compatible source of seed photons compatible is thermal synchrotron
emission of the hot plasma itself (which naturally produces homogenoeous and
isotropic seed photons). Then the soft excess found in the data can represent
either the peak of the self-absorbed thermal synchrotron emission (e.g.\
Wardzi\'nski \& Zdziarski 2000) or blackbody emission not related to the
dominant source of seed photons. The latter model was found in \S \ref{results}
to provide a good fit to the data. In fact, Wardzi\'nski \& Zdziarski (2000)
have predicted that the thermal synchrotron process is able to dominate the
supply of seed photons at an intermediate range of Eddington factors, $L_{\rm
X}/L_{\rm E}\la 10^{-3}$, compatible with the case in \source.

By considering the synchrotron process, we can constrain the magnetic field
strength in the source. The turnover energy, $E_{\rm t}$, at which the thermal
self-absorbed synchrotron emission peaks, is approximately 
\begin{equation}
\label{E_t} E_{\rm t}\simeq 1.3\! \left(kT\over 10^2{\rm keV}
\right)^{0.95}\!\! \tau^{0.05} \left(B\over 10^6\,{\rm G}\right)^{0.91}\! {\rm
eV} \end{equation} 
(Wardzi\'nski \& Zdziarski 2000). This yields a constraint on the magnetic 
field strength of $B\la 10^8$ G, with the approximate equality corresponding 
to 
the observed soft excess at $\sim 0.15$ keV (Fig.\ \ref{f:model}) being due to 
the turnover peak and no non-thermal electron tail. For lower values of $B$, 
the observed soft excess is due to blackbody emission. Then, the UV data of 
Hynes et al.\ (2000) constrain $E_{\rm t}\ga 10$ eV. Wardzi\'nski \& Zdziarski 
(2001) have found that $E_{\rm t}$ of equation (\ref{E_t}) can increase by a 
factor of $\sim 2$--3 in the presence of a weak non-thermal tail beyond the 
Maxwellian electron distribution (with a corresponding high-energy photon tail 
observed in Cyg X-1, McConnell et al.\ 2000). Then, the UV constraint implies 
$B\ga 10^7$ G in the absence of a non-thermal tail and  $B\ga 3\times 10^6$ G 
with a tail. Only lower values of the above range of $B$ correspond to 
equipartition in a hot accretion flow with moderately sub-virial ions [e.g., 
eq.\ (34) in Wardzi\'nski \& Zdziarski 2000], which indicates that the observed 
soft X-ray excess is due to a separate blackbody component rather than the 
turnover peak at $E_{\rm t}$. Summarizing, $B\la 10^8$ G is implied directly by 
the data, and  $B\ga (3$--$10)\times 10^6$ G for the sychrotron process to be 
the dominant source of seed photons.

The high erratic time variability we find in the soft X-ray band (as well as in
hard X-rays) points out to the origin of soft X-ray excess in the same spatial
region as that occupied by the hot plasma. This is in agreement with both the
seed photons being provided by the blackbody emission of clouds within the hot
plasma or their origin from the synchrotron process. The latter model also
agrees with a similar time variability observed in the optical band, as noted
by Merloni et al.\ (2000). 

Within the Compton scattering model in a hot central disk, our data appear 
compatible with the hot disk being advection-dominated, as recently proposed 
by Esin et al. (2001) on the basis of near--simultaneous {\it HST}, {\it EUVE},
{\it Chandra} and \xte\  observations. 
However, if such interpretation is correct, it is not clear how to compare 
this result with Cygnus X--1, which has a much larger luminosity and almost 
the same spectral parameters \cite{Esin98,Frontera01a}.

McClintock et al. (2001b) reported a break at about 2 keV in the $E F(E)$ 
spectrum of \source\/, which was interpreted by Esin et al. (2001) as due to 
the presence of a warm absorber. The data used by MCClintock et al. (2001b) 
were obtained combining together different instruments aboard different 
missions. We do not see this break (see Figure 8), that could also be due to 
cross--calibration problems.

\section{Conclusions}

Our main findings can be summarized as follows.

We find the Galactic column density of H{\sc i} measured in radio in the
direction to the source as $\nh\simeq (1.28$--$1.44)\times 10^{20}$ cm$^{-2}$
(Dickey \& Lockman 1990) to be consistent with spectral fitting of our \sax\/
data (see Table 1). On the other hand, we find the value of $2.8 \times
10^{20}$ cm$^{-2}$ favored by Dubus et al.\ (2001) based on their measurement
of Ca absorption to be intrinsically highly uncertain (\S \ref{nh}) as well
inconsistent with our data (\S \ref{results}).

Our \sax\/ spectrum of \source\/ shows a power law with $\Gamma\simeq 1.7$ 
modified by a soft X-ray excess and a high energy cutoff, with both features 
highly significant statistically. The soft excess contains a small fraction of 
the total flux, and is consistent with a blackbody component, either undergoing 
Comptonization or not. The high energy cutoff is well fitted by Comptonization 
in a thermal plasma with an electron temperature of $\sim 10^2$ keV and a 
Thomson optical depth of $\tau\sim 1$. In agreement with the weakness of the 
blackbody, Compton reflection is weak, also consistent with the weakness of an 
Fe K$\alpha$ line. The low amplitudes of all the features originating in the cold 
matter can have several explanations. If such components come from an underlying 
accretion disk, the relative large $\tau$ and inclination angle easily 
dim the observable reflection and blackbody features by a factor 
$\exp(-\tau/\cos i)\simeq 0.1$, consistent with our observations. However, 
in this case, the resulting Compton spectrum would be steeper than observed.  
A possibility is that any cold matter subtends a small solid angle as seen by the 
hot gas. 
In this case the seed photons for Comptonization can be provided by an outer 
cold disk, or by synchrotron emission internal in the hot disk, or both. 
If the geometry is that of a central hot disk, it is not clear if 
our data are consistent with the ADAF solution, or instead if they rather 
indicate a radiative cooled hot accretion disk.  

The data are statistically consistent with the reflector to have a low metallicity, 
$Z/Z_\sun\sim 0.1$, as expected due to the high latitute of the system. This could 
be an important finding, which should be tested by future observations of the 
secondary.

The values of $kT$ and $\tau$ we find are very similar to those typical of most
luminous (e.g., $L_{\rm X}\sim 0.02 L_{\rm E}$ in Cyg X-1) black-hole
binaries in the hard state, whereas the X-ray luminosity in our measurement is
only $\sim 10^{-3}L_{\rm E}$.

\acknowledgements

We thank Juri Poutanen for valuable discussions, and Grzegorz Wardzi\'nski
for reducing the \xte\ data used in \S \ref{comparison}. The \sax\/ program is
supported by the Italian Space Agency (ASI). AAZ and JM have been supported in
part by KBN grants 2P03D00614 and 2P03C00619p0(1,2) and a grant from the
Foundation for Polish Science. FF, FH and LS aknowledge a financial support 
from the Ministry of University and Scientific Research of Italy (COFIN 2000).


\begin{thebibliography}{}

{\small\parskip 0pt
%
\bibitem[Anders \& Ebihara 1982]{Anders82}
Anders, E., \& Ebihara, M. 1982, Geochim.\ Cosmochim.\ Acta, 46, 2363
%
\bibitem[Arnaud 1996]{Arnaud96}
Arnaud, K. A. 1996, in ASP Conf. Series 101, Astronomical Data
Analysis Software and Systems V, ed.\ G. H. Jacoby \& J. Barnes
(San Francisco: ASP), 17
%
\bibitem[]{bc96}
Baraffe, I., \& Chabrier, G. 1996, \apj, 461, L51
%
\bibitem[]{bk00}
Baraffe, I., \& Kolb, U. 2000, MNRAS, 318, 354
%
%\bibitem[Barret et al.\ 2000]{Barret00}
%Barret, D., Olive, J. F., Boirin, L., Done, C., Skinne, G. K., Grindlay, J. E.
%2000, ApJ, 533, 329
%
\bibitem[]{b99}
Beloborodov, A. M. 1999, \apj, 510, L123
%
%\bibitem[]{b99b}
%Beloborodov, A.  M. 1999b, in ASP Conf.\ Series 161, High Energy Processes in
%Accreting Black Holes, ed.\ J. Poutanen \& R. Svensson (San Francisco: ASP),
%295
%
\bibitem[]{bbkw98}
Beuermann, K., Baraffe, I., Kolb, U., \& Weichhold, M. 1998, A\&A, 339, 518
%
%\bibitem[Boella et al.\ 1997a]{Boella97a}
%Boella, G., et al.\ 1997a, \aaps, 122, 299
%
\bibitem[Boella et al.\ 1997]{Boella97}
Boella, G., et al.\ 1997, \aaps, 122, 327
%
\bibitem[Chaty et al.\ 2000]{Chaty00}
Chaty, S., et al.\ 2000, IAU Circ.\ 7394
%
\bibitem[Cook et al.\ 2000]{Cook00}
Cook, L., et  al. 2000, IAU Circ.\ 7397
%
\bibitem[Dickey \& Lockman 1990]{Dickey90}
Dickey, J. M., \& Lockman, F. J. 1990, ARA\&A 28, 215
%
\bibitem[]{dmz}
Done, C., Madejski, G. M., \& \.Zycki, P. T. 2000, \apj, 536, 213
%
\bibitem[Dubus et al.\ 2001]{Dubus00}
Dubus, G., Kim, R. S. J., Menou, K., Szkody, P., \& Bowen, D. V.  2001,
\apj, submitted (astro-ph/0009148)
%
\bibitem[Esin et al.\ 1998]{Esin98}
Esin, A. A., Narayan, R., Cui, W.,  Grove, J. E., \& Zhang, S. N. 1998,
\apj, 505, 854
%
\bibitem[Esin et al.\ 2001]{Esin01}
Esin, A. A., McClintock, J. E., Drake, J. J., Garcia M. R., Haswell, C. A., 
Hynes, R. I., \& Muno, M. P. 2001, ApJ, in press (astro-ph/0103044)
%
\bibitem[Faison et al. 1998]{Faison98}
Faison, M. D., Goss, W. M., Diamond, P. J., \& Taylor, G. B. 1998, AJ, 116,
2916
%
\bibitem[]{fkr92}
Frank, J., King, A., \& Raine, D. 1992, Accretion Power in Astrophysics
(Cambridge: Cambridge University Press)
%
\bibitem[Frontera et al.\ 1997]{Frontera97}
Frontera, F., et al.\ 1997, \aaps, 122, 357
%
\bibitem[Frontera et al.\ 2001a]{Frontera01a}
Frontera, F., et al.\ 2001a, \apj, 546, 1027
%
\bibitem[Frontera et al. 2001b]{Frontera01b}
Frontera, F., et al.\ 2001b, Proc. 4th INTEGRAL Workshop, in press.
%
\bibitem[Garcia et al.\ 2000]{Garcia00}
Garcia, M., et~al.\ 2000, IAU Circ.\ 7392
%
\bibitem[]{g00}
Garnavich, P. M. 2000, IAU Circ.\ 7542
%
%\bibitem[]{gf91}
%George, I. M., \& Fabian, A. C. 1991, \mnras, 249, 352
%
\bibitem[]{g97}
Gierli\'nski, M., Zdziarski, A.  A., Done, C., Johnson, W.  N., Ebisawa,
K., Ueda, Y., Haardt, F., \& Phlips, B.  F. 1997, \mnras, 288, 958
%
\bibitem[]{g99}
Gierli\'nski, M., Zdziarski, A.  A., Poutanen, J., Coppi, P.,
Ebisawa, K., \& Johnson W.  N. 1999, \mnras, 309, 496
%
%\bibitem[Gilfanov, Churazov \& Revnivtsev  1999]{Gilfanov99}
%Gilfanov, M., Churazov, E., \& Revnivtsev, M. 1999, \aap, 352, 182
%
%\bibitem[]{h90}
%Hall, D. S. 1990, AJ, 100, 554
Haardt, F., \& Maraschi, L. 1991, ApJ, 380, L51
%
\bibitem[Haswell, Hynes, \& King 2000a]{Haswell00a}
Haswell, C. A., Hynes, R. I., \& King, A. R. 2000a, IAU Circ.\ 7407
%
\bibitem[Haswell et~al.\ 2000b]{Haswell00b}
Haswell, C. A., et~al.\ 2000b, IAU Circ.\ 7427
%
\bibitem[Hynes et al.\ 2000]{Hynes00}
Hynes, R. I., Mauche C. W., Haswell, C. A., Shrader, C. R., Cui, W., \& Chaty,
S. 2000, \apj, 539, L37
%
\bibitem[Kuulkers 2001]{Kuulkers01}
Kuulkers, E. 2001, Astr. Nachr. 322, 9 (astro-ph/0102066)
%
\bibitem[]{mz95}
Magdziarz, P., \& Zdziarski, A.  A. 1995, \mnras, 273, 837
%
\bibitem[]{m01}
Malzac, J., Beloborodov, A. M., \& Poutanen, J. 2001, \mnras, in press

\bibitem[Manzo et al.\ 1997]{Manzo97}
Manzo, G., et al.\ 1997, \aaps, 122, 341
%
\bibitem[]{mc00}
McClintock, J., Garcia, M., Zhao, P., Caldwell, N., \& Falco, E. 2000, IAU
Circ.\ 7542
%
\bibitem[]{mc01a}
McClintock, J. E., Garcia, M. R., Caldwell, N., Falco, E. E., Garnavich, P. M.,
\& Zhao, P. 2001a, \apj, submitted (astro-ph/0101421) [M01a]
%
\bibitem[McClintock et al. 2001a]{McClintock01b}
McClintock, J. E., et al. 2001b, \apj, submitted (astro-ph/0103051)
%
\bibitem[]{mcc00}
McConnell, M. L., et al.\ 2000, ApJ, 543, 928
%
\bibitem[Merloni, Di Matteo, \& Fabian 2000]{Merloni00}
Merloni, A., Di Matteo, T., \& Fabian, A. C. 2000, \mnras, 318, L15
%
\bibitem[Morrison \& McCammon 1983]{Morrison83}
Morrison, R., \& McCammon, D. 1983, \apj, 270, 119
%
\bibitem[]{ny95}
Narayan, R., \& Yi, I. 1995, ApJ, 452, 710
%
%\bibitem[]{n00}
%Nayakshin, S., Kazanas, D., \& Kallman, T. R. 2000, \apj, 537, 833
%
\bibitem[Parmar et al.\ 1997]{Parmar97}
Parmar, A. N., et al.\ 1997, \aaps, 122, 309
%
\bibitem[Pooley \& Waldram 2000]{Pooley00}
Pooley, G. G., \& Waldram, E. M. 2000, IAU Circ.\ 7390
%
\bibitem[Poutanen \& Svensson 1996]{Poutanen96}
Poutanen, J., \& Svensson, R. 1996, ApJ, 470, 249
%
\bibitem[]{p76}
Pringle J. E., 1976, MNRAS, 177, 65
%
\bibitem[Remillard et al.\ 2000]{Remillard00}
Remillard, R. A., et~al.\ 2000, IAU Circ.\ 7389
%
\bibitem[Revnivtsev, Sunyaev, \& Borozdin 2000]{Revnivtsev00}
Revnivtsev, M., Sunyaev, R., \& Borozdin, K. 2000, A\&A, 361, L37
%
\bibitem[Schlegel, Finkbeiner \& Davis]{Schlegel98}
Schlegel, D.J.,  Finkbeiner, D.P. \& Davis, M. 1998, ApJ, 500, 525
%
\bibitem[]{sd94}
Sembach, K. R., \& Danks, A. C. 1994, \aap, 289, 539
%
\%bibitem[]{ss73}
%Shakura, N. I., \& Sunyaev, R. A. 1973, \aap, 24, 337
%
\bibitem[]{sle}
Shapiro, S. L., Lightman, A. P., \& Eardley, D. M. 1976, ApJ, 204, 187
%
\bibitem[]{s81}
Smak, J. 1981, Acta Astr., 31, 395
%
%\bibitem[Sunyaev \& Revnivtsev 2000]{Sunyaev00}
%Sunyaev, R., \& Revnivtsev, M. 2000, A\&A, 358, 617
%
%\bibitem[Sunyaev \& Titarchuk 1980]{Sunyaev80}
%Sunyaev, R., \& Titarchuk, L. G. 1980, A\&A, 86, 121
%
%\bibitem[Stull, Ioannou, \& Webb 2000]{Stull00}
%Stull, J., Ioannou, Z., \& Webb, N. A. 2000, IAU Circ.\ 7407
%
\bibitem[]{sz94}
Svensson, R., \& Zdziarski, A. A. 1994, ApJ, 436, 599
%
\bibitem[Tanaka \& Shibazaki 1996]{Tanaka96}
Tanaka, Y., \& Shibazaki, N. 1996, ARA\&A, 34, 607
%
\bibitem[Titarchuk 1994]{Titarchuk94}
Titarchuk, L. G. 1994, \apj, 434, 570
%
%\bibitem[Uemura et al.\ 2000a]{Uemura00a}
%Uemura, M. et~al.\ 2000a, IAU Circ.\ 7418
%
\bibitem[Uemura et al.\ 2000]{Uemura00}
Uemura, M., et~al.\ 2000, PASJ, 52, L15
%
\bibitem[van Paradijs \& McClintock 1995]{Jvp95}
van Paradijs, J., \& McClintock, J. E. 1995, in X-ray Binaries, ed.\ W. H. G.
Lewin, J. van Paradijs \& E. P. J. van den Heuvel (Cambridge: Cambridge
University Press), 58
%
\bibitem[]{wg00}
Wagner, R. M., Foltz, C. B., Starrfield, S. G., \& Hewett, P. 2000, IAU Circ.\
7542
%
\bibitem[]{Wardzinski00}
Wardzi\'nski, G., \& Zdziarski, A. A. 2000, \mnras, 314, 183
%
\bibitem[]{Wardzinski01}
Wardzi\'nski, G., \& Zdziarski, A. A. 2001, \mnras, in press
%
%\bibitem[]{wlz}
%White, T. R., Lightman, A. P., \& Zdziarski, A. A. 1988, \apj, 331, 939
%
\bibitem[]{w90}
Wilson, R. E. 1990, ApJ, 356, 615
%
\bibitem[Wilson \& McCollough 2000]{Wilson00}
Wilson, C. A., \& McCollough, M. L. 2000, IAU Circ.\ 7390
%
\bibitem[Wood et al.\ 2000]{wd00}
Wood, K. et al.\ 2000, ApJ, 544, L45
%
\bibitem[Wren \& McKay 2000]{Wren00}
Wren, J., \& McKay, T. 2000, IAU Circ.\ 7394
%
%\bibitem[]{z98}
%Zdziarski, A. A. 1998, \mnras, 296, L51
%
\bibitem[]{z00}
Zdziarski, A. A. 2000, in IAU Symp.\ 195, Highly Energetic Physical Processes
and Mechanisms for Emission from Astrophysical Plasmas, ed.\ P. C. H. Martens,
S. Tsuruta \& M. A. Weber (San Francisco: ASP), 153 (astro-ph/0001078)
%
\bibitem[]{zjm96}
Zdziarski, A. A., Johnson, W. N., \& Magdziarz, P. 1996, MNRAS, 283, 193
%
}
\end{thebibliography}
\end{document}